# Large memcapacitance and memristance at Nb:SrTiO$_3$/La$_{0.5}$Sr$_{0.5}$Mn$_{0.5}$Co$_{0.5}$O$_{3-\delta}$ Topotactic Redox Interface

W. R. Acevedo[1,2], C. A. M. van den Bosch[3], M. H. Aguirre[4,5,6], C. Acha[2,7], A. Cavallaro[3], C. Ferreyra[1,2], M. J. Sánchez[2,8], L. Patrone[9], A. Aguadero[3,*], D. Rubi[1,2,*]

[1]*Comisión Nacional de Energía Atómica and Instituto de Nanociencia y Nanotecnología, Centro Atómico Constituyentes, 1650, Buenos Aires, Argentina*

[2] *Consejo Nacional de Investigaciones Científicas y Técnicas, Godoy Cruz 2290 (1425), Buenos Aires, Argentina.*

[3] *Department of Materials, Imperial College London, London, SW7 2AZ, United Kingdom*

[4]*Departmento de Física de Materia Condensada, Universidad de Zaragoza, Pedro Cerbuna 12  50009 Zaragoza - Spain*

[5] *Laboratorio de Microscopías Avanzada (LMA), Instituto de Nanociencia de Aragón (INA)-Universidad de Zaragoza,C/Mariano Esquillor s/n. 50018 Zaragoza, Spain*

[6] *Instituto de Ciencias de Materiales de Aragón (ICMA), Universidad de Zaragoza, Zaragoza, Spain*

[7] *Depto. de Física, FCEyN, Universidad de Buenos Aires & IFIBA, UBA-CONICET, Pab I, Ciudad Universitaria, Buenos Aires (1428), Argentina*

[8] *INN, Centro Atómico Bariloche and Instituto Balseiro, 8400 San Carlos de Bariloche, Argentina*

[9] *INTI, CMNB, Av. Gral Paz 5445, B1650KNA San Martín, Buenos Aires, Argentina*

The possibility to develop neuromorphic computing devices able to mimic the extraordinary data processing capabilities of biological systems spurs the research on memristive systems. Memristors with additional functionalities such as robust memcapacitance can outperform standard devices in key aspects such as power consumption or miniaturization possibilities. In this work, we demonstrate a large memcapacitive response of a perovskite memristive interface, using the topotactic redox ability of La$_{0.5}$Sr$_{0.5}$Mn$_{0.5}$Co$_{0.5}$O$_{3-\delta}$ (LSMCO, $0 \leq \delta \leq 0.62$). We demonstrate that the multi-mem behaviour originates at the switchable n-p diode formed at the Nb:SrTiO$_3$/LSMCO interface. We found for our Nb:SrTiO$_3$/LSMCO/Pt devices a memcapacitive effect $C_{HIGH}/C_{LOW} \sim$ 100 at 150kHz. The proof-of-concept interface reported here opens a promising venue to use topotactic redox materials for disruptive nanoelectronics, with straightforward applications in neuromorphic computing technology.

\* Corresponding authors: diego.rubi@gmail.com (DR), a.aguadero@imperial.ac.uk (AA)

Neuromorphic computing devices aim mimicking biological systems and are expected to dramatically improve performance and efficiency of electronic devices for advanced information technology.[1] Brain synapses can be emulated by memristors,[2,3] consisting of capacitor-like structures displaying a reversible and non-volatile electrical resistance change upon the application of electrical stimulus.[4,5] Other potential applications of memristors include nanoelectronic memories [2] and logic gates.[6] Memristive behavior is ubiquitously found in transition metal oxides, including perovskite manganites.[7] Proposed memristive mechanisms for metal/manganite systems include the modulation of metal/insulator Schottky barriers due to oxygen vacancies (OV) electromigration, [8-10] or the interfacial redox reaction occurring when a reactive electrode (Ti or Al) is used.[11-13] In these cases, oxygen exchange with the environment is neglected.[9] Alternatively, reports on memristive perovskites claiming for oxygen exchange between the perovskite and the atmosphere [14] or the metallic electrode [15] can be found. For volumetric redox processes triggered by electrical stimuli, a robust memristive effect is expected for perovskites displaying topotactic redox ability, i.e. the capability of reversibly storing and releasing oxygen with slight structural changes that maintain the perovskite structure. [16,17] Here, the memristive effect relies on the electrical switch between oxidized and reduced phases with different electrical conductivities. A better stability for redox memristive behavior is envisaged for topotactic perovskites in comparison with standard ones, [18,19] as the structural changes –i.e. change in the perovskite space group- [17] in the former case allows transitions between structures presenting well defined minimum energies.

Memcapacitance –non-volatile change of a device capacitance C upon the application of electrical stress- is an additional functionality of memristors that has scarcely been explored.[20-26] Proposed mechanisms include creation/annihilation of conducting nanofilaments, [20,21] modulation of Schottky barriers at interfaces,[21,22] oxidation/reduction of a $TiO_x$ active layer [23] or changes in the oxide permittivity upon OV electromigration.[24,25] While applications for memcapacitance, including neuromorphic computing devices, have been proposed, [27] the interest in this phenomenon has been hampered by the small reported figures to date ($C_{HIGH}/C_{LOW} \leq 10$).[20-26] It was shown that associated capacitive networks, suitable for efficient pattern recognition, can be build from cells able to switch their capacitance between $C_{HIGH}$ and $C_{LOW}$, where the array size –linked to the device computing capability- scales with $C_{HIGH}/C_{LOW}$ ratio.[28] This evidences the high technological interest of memcapacitive systems with large response.

In this paper we show that the interface between the topotactic redox perovskite manganite $La_{0.5}Sr_{0.5}Mn_{0.5}Co_{0.5}O_{3-\delta}$ (LSMCO, $0 \leq \delta \leq 0.62$, p-type)[17] and $Nb:SrTiO_3$ (NSTO, n-type) behaves as a switchable n-p diode with memristive and large memcapacitive behavior. We show that the observed multi-mem behavior is related to the electrical switch between LSMCO oxidized ($\delta = 0$) and reduced ($\delta \approx 0.62$) phases.

Oxidized phase LSMCO thin films were grown by laser ablation on Nb:SrTiO$_3$ (0.5 wt%, 001) substrates. The growth temperature, oxygen pressure and laser fluence were fixed at 800 ºC, 0.04 mbar and 0.5 – 1 J/cm$^2$ respectively. Pristine LSMCO films were ≈ 20 nm thick and epitaxial, as shown in Suppl. Figure S1. High resolution scanning transmission electron microscopy (STEM) was performed using a FEI Titan G2 microscope with probe corrector. Microstructured top Pt electrodes, ~20 nm thick, were fabricated by FIB or optical lithography. The platinum high work function (5.6 eV) allows an ohmic interface with p-type LSMCO. [29] The NSTO substrate was grounded (a drop of silver paint was used to make contact) and the electrical stimulus was applied to the top Pt electrode. For the electrical characterization we used a Keithley 2612 source-meter, an AutoLab PGSTAT302N impedance analyzer and a standard LCR-meter.

The virgin resistances of 37.5 x 10$^3$ µm$^2$ NSTO/LSMCO/Pt devices were ~1 MΩ, and the current-voltage (I-V) curve for low stimulus, shown in Suppl. Figure S2, displays a rectifying behavior linked to the formation of an n-p diode at the NSTO / LSMCO interface. A forming process is triggered when a -7 V pulse is applied, resulting in a sudden resistance drop to ~50 Ω (Suppl. Figure S2). After forming, dynamic I-V curves were obtained by applying a sequence of 1ms voltage pulses of different amplitudes (0→V$_{MAX}$→V$_{MIN}$→0), with the current measured during the application of the pulse. Additionally, after each voltage pulse a small reading voltage (100mV) was applied to determine the remnant resistance states, obtaining the hysteresis switching loops (HSLs). Figures 1(a) and (b) display typical I-V and HSL curves, both demonstrating the memristive properties of our devices. The device switches from a low resistance (R$_{LOW}$) to high resistive state (R$_{HIGH}$) (RESET process) upon the application of ~ +5.5 V, while the opposite behavior (SET process) is observed upon the application of ~ –1.5 V. From the second cycle, R$_{HIGH}$ and R$_{LOW}$ stabilize to ~ 2-6 kΩ and ~ 100-200 Ω, respectively, giving an average ON/OFF ratio of ~ 25. Figure 1(c) shows an endurance test with a stable behavior for ~ 200 cycles, while Figure 1(d) shows retention times of at least 10$^4$ s for both resistive states.

Interestingly, the observed memresistance is concomitant with a large memcapacitive effect. Dynamic capacitance-voltage (C-V) curves were obtained by measuring the device capacitance upon the application of DC voltage pulses of increasing amplitude with a small superimposed AC signal (10 kHz, amplitude 200 mV). Figure 2(a) displays dynamic C-V curves measured on devices prepared in both R$_{HIGH}$ and R$_{LOW}$ states. An evident difference between both curves is observed, indicating a significant capacitance change between $R_{HIGH}$ and $R_{LOW}$ states, where the $R_{HIGH}$ ($R_{LOW}$) state corresponds to a low (high) capacitance $C_{LOW}$ ($C_{HIGH}$) one. The negative capacitance found for $C_{LOW}$ for positive voltages is attributed to the non-monotonic or positive-valued behaviour of the time-derivative of the transient current in response to a small voltage step.[30] The memcapacitive effect was confirmed by remnant capacitance measurements, performed in both states by applying a pure AC signal at different frequencies. Figure 2(b) displays the evolution of both remnant $C_{LOW}$ and $C_{HIGH}$ states as a function of the frequency (f) of the excitation signal. $C_{LOW}$ ~ 3.5 pF at 10kHz and displays a subtle decrease with f, while $C_{HIGH}$ displays a stronger decrease, from ~ 3 nF at 10 kHz to ~ 0.13 nF at 300 kHz. The capacitance decrease is at a higher rate for f < 100 kHz. The existence of leakage channels at the diode interface likely increase its effective capacitance at low f, [31] but other effects such as the presence of surface states might also contribute to the dependence of the capacitance with f. [32] In the low frequency range there is a rapid drop in the $C_{HIGH}/C_{LOW}$ ratio, from ~ 900 at 10 kHz to ~ 130 at ~ 100 kHz, likely due to leakage effects. For higher frequencies leakage effects should not contribute [31] and the dependence of $C_{HIGH}/C_{LOW}$ with f is milder. We obtained $C_{HIGH}/C_{LOW}$ ~ 100 at ~ 150 kHz, which is around one order of magnitude larger than memcapacitive figures reported to date for other systems [20-26]. The ability of our devices to reversibly change their capacitance between two non-volatile states upon consecutive cycling is confirmed by Figure 2(c), while Figure 2(d) displays retention times for both $C_{HIGH}$ and $C_{LOW}$ higher than $10^3$ s, respectively.

Figure 3(a) shows a scanning electron microscopy top-view of device after forming, where three distinct zones are identified. Zone 1 has a diameter of ~5 μm and corresponds to the contact position of the tip. Zone 2 is a ring of higher contrast than Zone 1 (diameter: ~15 μm) where material from the film and the Pt electrode has been expelled during the forming process probably due to the release of $O_2$ gas. [33]. At Zone 1, STEM-HAADF analysis shows that, upon forming, the pristine LSMCO epitaxial nanostructure (shown in Suppl. Figure S3(a)) re-crystallizes due to self heating effects. This re-crystallization can comprise the complete LSMCO thickness, leading to an arrangement of nanograins as the one displayed in Figures 3(b) and S3(c). The bottom LSMCO nanograins, in contact with the NSTO substrate, retain a (001) out-of-plane orientation, while the top grains, in contact with the Pt top electrode, are in general not coherent with the bottom grains and present a tilted (001) direction, as shown in the zoomed image of Figure 3(c). LSMCO re-crystallization at Zone 1 can also be partial, leading to top LSMCO non-coherent nanograins located on top of epitaxial LSCMO (see Figure S3(c)). LSMCO re-crystallization process is likely driven by to the presence of thermal self-accelerated effects [34] where both the Pt electrode (see Suppl. Figure S3(b),(c)) and (part of) LSMCO melt during electroforming, followed by a fast cooling after the end of the forming voltage that quenches melted LSMCO into a non-coherent nanograins arrangement. TEM analysis also shows that part of LSMCO becomes reduced upon for forming. Figure 4 shows atomic images and Fast Fourier Transforms corresponding to two nanograins at Zone 1. The nanograin displayed in Figure 4(a) and (b), located close to the LSMCO/NSTO interface, remains structurally very similar to pristine (oxidized) LSMCO, while the one displayed in Figure 4(c) and (d), located close to the Pt top electrode, is reduced and presents an ordered structure of OV that double the LSMCO unit cell along the (001) direction, resembling previous reports in brownmillerite.[35] These grains are likely the source of the released $O_2$ during forming. Zone 3 displays a similar contrast as Zone 1 and presents an epitaxial structure with the presence of extended defects, as analyzed in Suppl. Figure S4. All remnant resistive and capacitive states are independent of the (virgin) device area (Suppl. Figure S5), indicating that the forming process electrically decouples Zone 1 from the rest of the device. Thus, the multi-mem behavior is confined to Zone 1, presenting an effective area of ~ 80 $\mu m^2$.

We propose that the LSMCO multi-mem behavior is related to its topotactic redox ability where oxidation (reduction) of LSMCO nanograins in Zone 1 is associated with the SET (RESET) process. This scenario is supported by memresistance experiments performed in vacuum (<1x10$^{-2}$ mbar) where it is found that the SET event is not achieved (Suppl. Figure S6), indicating that environmental $O_2$ is critical for the SET process through LSMCO oxidation. Further evidence about the link between memresistance and LSMCO redox was obtained by simulating the experimental HSL (Figure 1(b)) with the voltage enhance OV drift model, [8,9,13] adapted to the present system. The simulation assumes a 1D chain of LSMCO nanodomains, able to accommodate different oxygen content which controls their resistivity, in contact with an oxygen reservoir (see Suppl. Mater. for further details). The model simulates the oxygen dynamics related to the electrically induced LSMCO oxidation and reduction. The simulated HSL shows a very good agreement with the experimental one, as shown in Figure 1(b).

Further insight into the memristive mechanism was obtained from the analysis of the I-V curves corresponding to the two resistive states, by plotting the power exponent $\gamma = d(\ln(I))/d\ln(V)$ vs. $V^{1/2}$.[36] This method is useful for identifying the presence of multiple conduction mechanisms,[37-39] which often occurs in metal/complex oxide interfaces.[40-42] Figure 5(a) shows the complex evolution of $\gamma$ vs. $V^{1/2}$ for both $R_{HIGH}$ ($C_{LOW}$) and $R_{LOW}$ ($C_{HIGH}$), indicating the presence of several circuit elements with relative weights that change between these states. The equivalent circuit that describes the $\gamma$ vs. $V^{1/2}$ behavior is shown in Figure 4(e) and includes the series combination of: i) an n-p diode in parallel with a leakage channel $R_1$, corresponding to the NSTO/ LSMCO interface, ii) the series resistor $R_2$ associated with the ohmic conduction of non-interfacial LSMCO plus the LSMCO/Pt interface, and iii) a Schottky diode linked to the external Ag/NSTO contact. The experimental I-V curves were fitted by numerically solving the implicit I-V equations of the circuit (see Suppl. Mater.), and all circuit parameters were extracted and listed in Suppl. Table S1. Figures 5(a) and (b) shows the excellent fits of the experimental and $\gamma$ vs. $V^{1/2}$ and I-V curves for both $R_{HIGH}$ ($C_{LOW}$) and $R_{LOW}$ ($C_{HIGH}$) states. The fittings indicate that the transition between $R_{HIGH}$ ($C_{LOW}$) and $R_{LOW}$ ($C_{HIGH}$) is dominated by the metallization of the LSMCO/NSTO interface, reflected in the increase of the p-n diode inverse saturation $I_{satpn}$ from 2.5 to 5.6 µA and the decrease of the leakage resistance $R_1$ from 270 to 110 Ω. Impedance spectroscopy (IS) was performed to further investigate the AC response of the different multi-mem states. Figures 5(c) and (d) display the Cole-Cole plots for $R_{HIGH}$ ($C_{LOW}$) and $R_{LOW}$ ($C_{HIGH}$) respectively. The equivalent circuit that allows a good fit of the $R_{LOW}$ ($C_{HIGH}$) state is shown in Figure 5(f), and was also used to simulate the $R_{HIGH}$ ($C_{LOW}$) spectrum. The fitted values for the circuit elements are shown in Suppl. Table S2. Again, it is found that the multi-mem effect is mainly localized at the NSTO/LCMO interface, characterized by the parallel combination of $R^*_1$ and $C^*_1$ (see Figure 5(f)). $R^*_1$ changes from ~10 Ω to ~500 Ω and $C^*_1$ from ~ $3\times10^{-8}$ F to ~ $1\times10^{-12}$ F between $R_{LOW}$ ($C_{HIGH}$) and $R_{HIGH}$ ($C_{LOW}$), reinforcing the idea of the presence of a switchable diode formed at the NSTO/LSMCO interface.

Thus, an oxidized NSTO/LSMCO interface leads to a $R_{LOW}$ ($C_{HIGH}$) state, and a reduced interface results in a $R_{HIGH}$ ($C_{LOW}$) state. The key factor for this behavior is the topotactic redox ability of LSMCO, which tolerates large changes in its oxygen content between oxidized and reduced phases. The physical origin of the large memcapacitance is intriguing and can be attributed to different effects. A possible one is related to large variations in the donors/acceptors balance at the NSTO/LSMCO interface upon LSMCO redox. The p-character of oxidized LSMCO is determined by the 0.5 holes/f.u. introduced in the lattice when $Sr^{2+}$ ions replace $La^{3+}$. In the reduced phase, each OV leaves two electrons behind, and part of these electrons will recombine with existing holes, reducing the number of uncompensated LSMCO acceptors. This could strongly change the (NSTO) donors / (LSMCO) acceptors balance at the interface, affecting the diode depletion layer and its capacitance. Other possible origin for the large memcapacitance is the Maxwell-Wagner effect, related to the creation of metallic zones embedded in the dielectric oxide, acting as the parallel plates of nanocapacitors, [43] that could produce large effective dielectric constants. [44] The metallization of the NSTO/LSMCO interface in the $R_{LOW}$ ($C_{HIGH}$) state could take place inhomogeneously, generating an interfacial capacitance significantly larger than that expected for a homogeneous interface. Further studies are necessary to confirm and get deeper insight into these mechanisms, together with the development of strategies to circumvent the electroforming process.

See Suplementary Material for additional structural and electrical characterization of our devices, together with details about the simulation and fittings of the electrical response.

D.R. acknowledges financial support from ANPCyT, projects PICTs 0867 and 1836. Helpful discussions with S. Menzel, P. Stoliar, M. Rozenberg V. Ferrari and P. Levy are acknowledged. We thank F. Golmar, from INTI, for the access to the FIB facility. A.A., C.vdB. and A.C. acknowledge the support of the Engineering and Physical Sciences Research Council (EPSRC), Grants No. EP/M014142/1, EP/P026478/1 and EP/L504786/1. A.A. and A. C. also acknowledge FETPROACT-2018-2020 "HARVESTORE" 824072 project. M.A. acknowledges financial support of H2020-MSCA-RISE-2016 SPICOLOST Grant No. 734187 to perform TEM studies at LMA-INA, University of Zaragoza.

Figure Captions

Figure 1: (a) Dynamic pulsed I-V curve recorded on a formed device, with arrows indicating the circulation direction of the curve. The inset displays the I-V curve corresponding to a non-formed device; (b) Experimental hysteresis switching loop (HSL, symbols), recorded simultaneously with the I-V curve. The simulated HSL is displayed with a solid line; (c) Retention experiments corresponding to a NSTO/LSMCO/Pt device, for both $R_{LOW}$ and $R_{HIGH}$ states; (d) Endurance test performed by applying single SET and RESET voltage pulses with opposite polarities.

Figure 2: (a) Dynamic capacitance-voltage (C-V) curves recorded both for oxidized ($R_{LOW}$ and $C_{HIGH}$) and reduced ($R_{HIGH}$ and $C_{LOW}$) LSMCO states; (b) Evolution of remnant $C_{HIGH}$ and $C_{LOW}$ states with the frequency of the external AC signal; (c) Reversible switch between $C_{HIGH}$ and $C_{LOW}$ (measured at 10kHz) after the application of SET and RESET single pulses. $C_{HIGH}$ and $C_{LOW}$ values were in the ranges 3-20 nF and 3-9 pF, respectively; (d) Retention experiments for $C_{HIGH}$ and $C_{LOW}$ states, measured at 10 kHz.

Figure 3: (a) Scanning electron microscopy image (top-view) of a formed device. Three zones are identified and described in the text; (b) STEM-HAADF cross-section (Zone 1) corresponding to a formed device; (c) Higher magnification STEM-HAADF cross-section evidencing the non-coherent nature of LSMCO nanograins at Zone 1 after forming. Atomic planes are indicated with green lines and the grain boundary is marked with white dotted line; (d) Sketch of the forming, RESET and SET processes in the NSTO/LSMCO/Pt devices. After forming, epitaxial LSMCO is re-crystallized and nanograins are formed. The transition between $R_{LOW}$ ($C_{HIGH}$) and $R_{HIGH}$ ($C_{LOW}$) is related to LSMCO oxidation and reduction.

Figure 4: (a), (b) Atomic image and FFT, respectively, corresponding to a post-forming LSMCO nanograin (Zone 1) in contact with the NSTO substrate. An oxidized LSMCO perovskite structure is observed; (c), (e) Same analysis for a LSMCO nanograin in contact with the Pt top electrode. The grain structure is perovskite-type, but with ordered OV that double the unit cell along the (001) direction (see additional diffraction spots, circled in red), indicating the presence of reduced LSMCO. The zone axis is [-1,10] for both FFTs. These images correspond to a different lamella than images displayed in Figure 3.

Figure 5: (a) γ vs. $V^{1/2}$ representation and (b) corresponding I-V curves for both the experimental (open symbols) and the calculated (solid line) $R_{HIGH}$ ($C_{LOW}$) and $R_{LOW}$ ($C_{HIGH}$) states. The fits are performed by considering the (DC) equivalent circuit presented in (e). $C_1$ and $C_2$ were included for completeness but have no effect at the low frequency range used for the I-V experiments; (c), (d) Impedance spectroscopy spectra recorded for $R_{HIGH}$ ($C_{LOW}$) and $R_{LOW}$ ($C_{HIGH}$) states. The experimental points are shown with open symbols, while the fittings are shown with solid lines; (f) Equivalent (AC) circuit proposed to model the experimental impedance spectra. $R^*_3//C^*_3$ corresponds to LSMCO/Pt interface, $R^*_1//C^*_1$ to NSTO/LSMCO interface, and the series resistance $R^*_2$ includes the non-interfacial LSMCO plus the contribution of the Ag/NSTO interface. We note that $R_2$ in (e) includes both $R^*_2$ and $R^*_3$ of the AC equivalent circuit of (f) and that the n-p diode AC contribution is included in $R_1^*$.

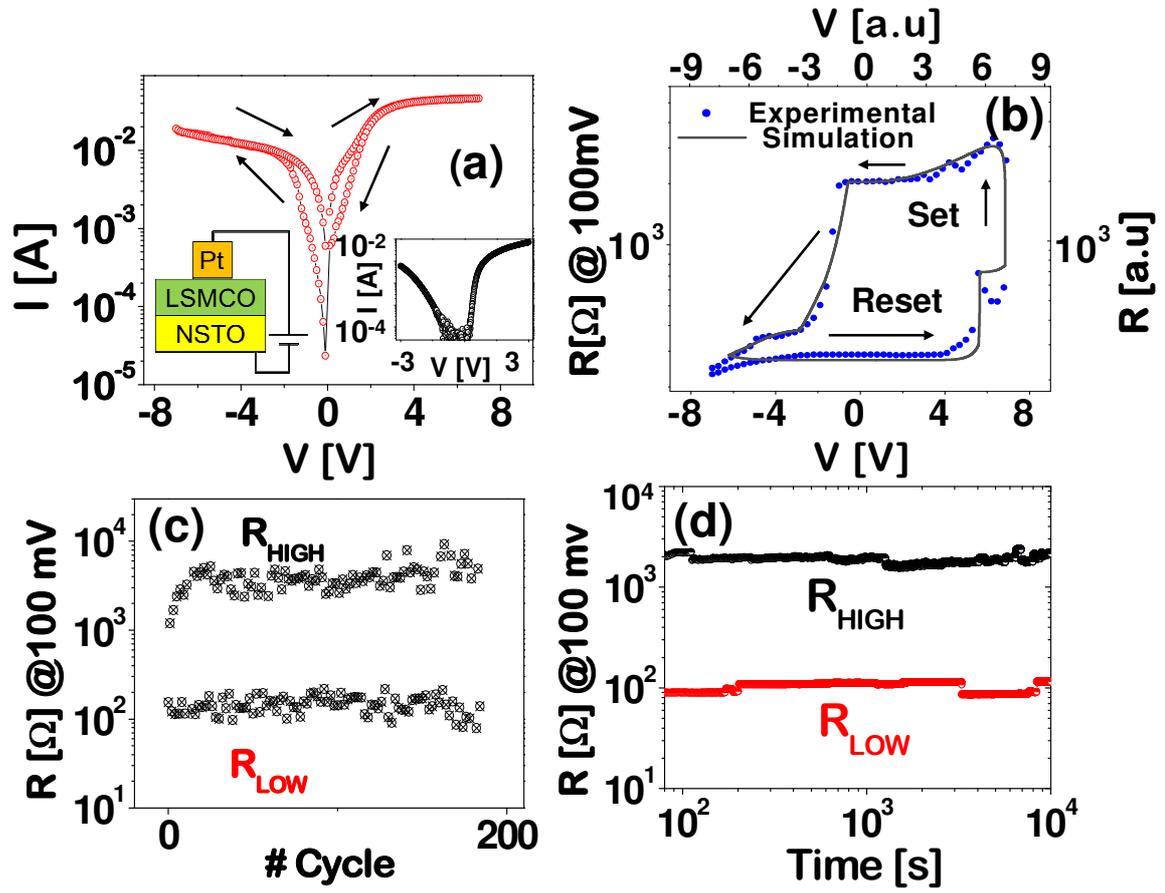

**FIGURE 1**

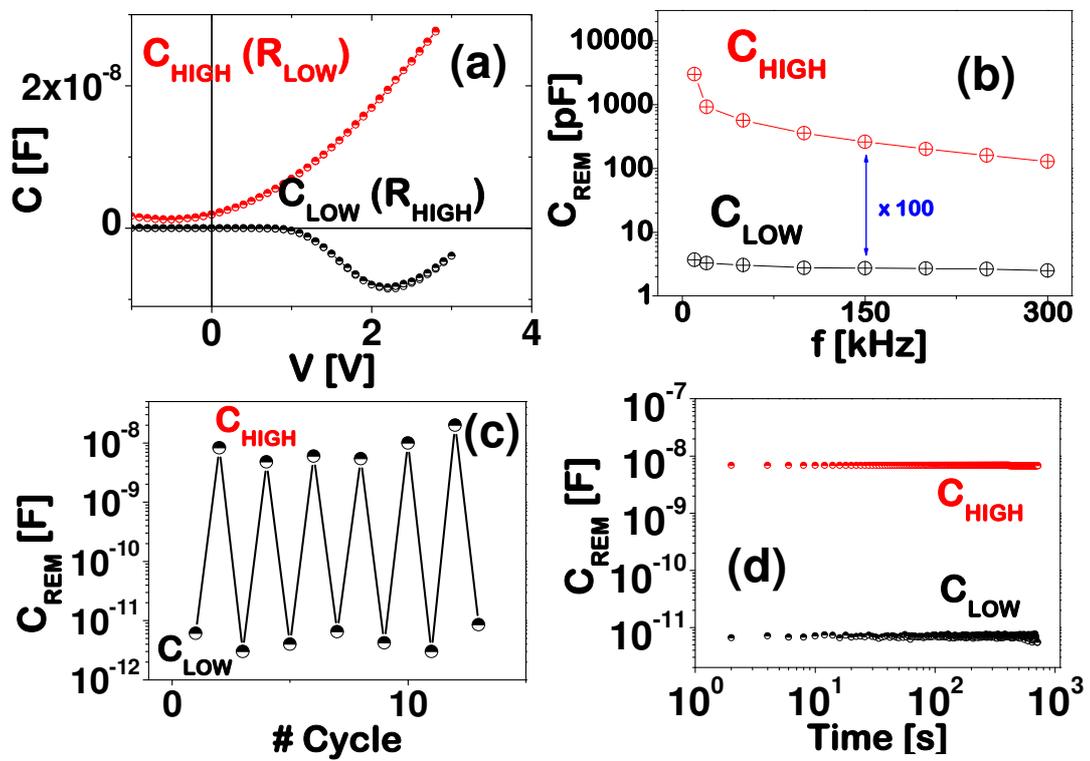

**FIGURE 2**

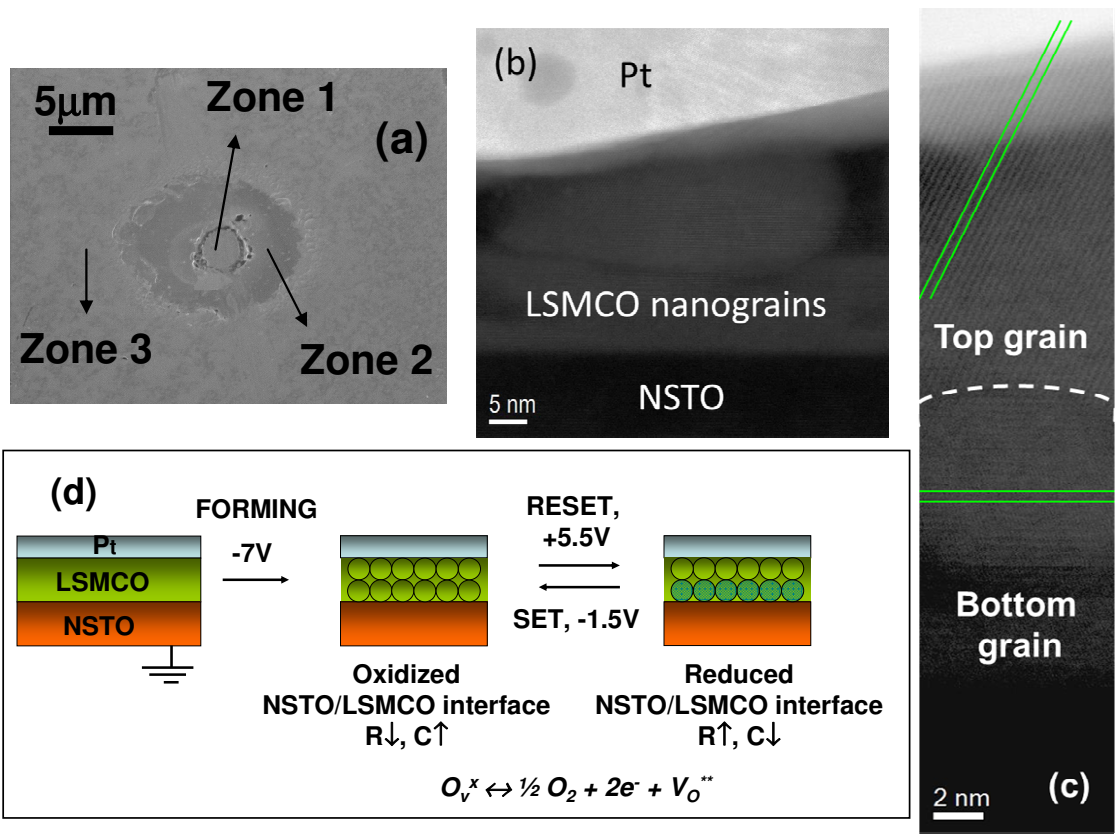

**FIGURE 3**

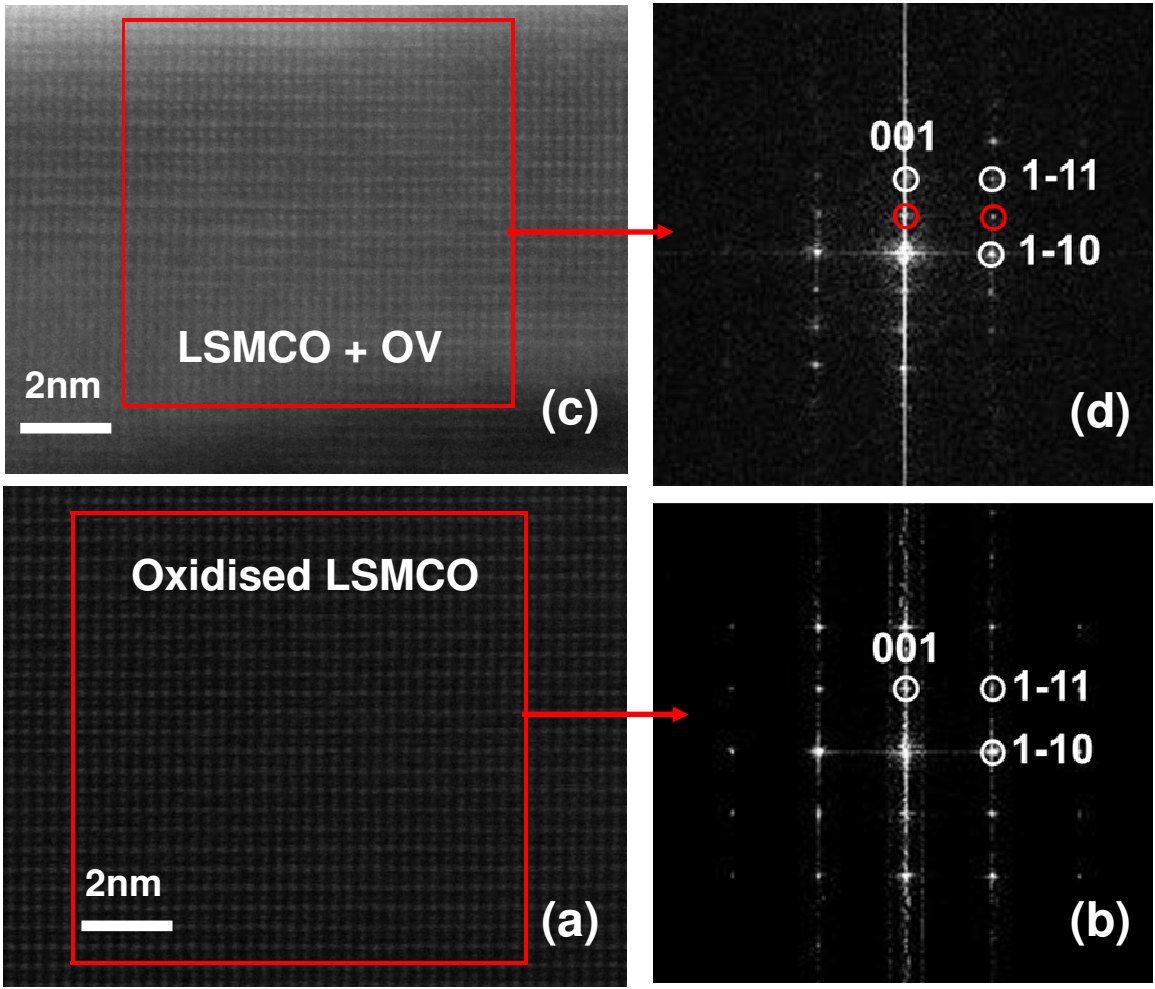

**FIGURE 4**

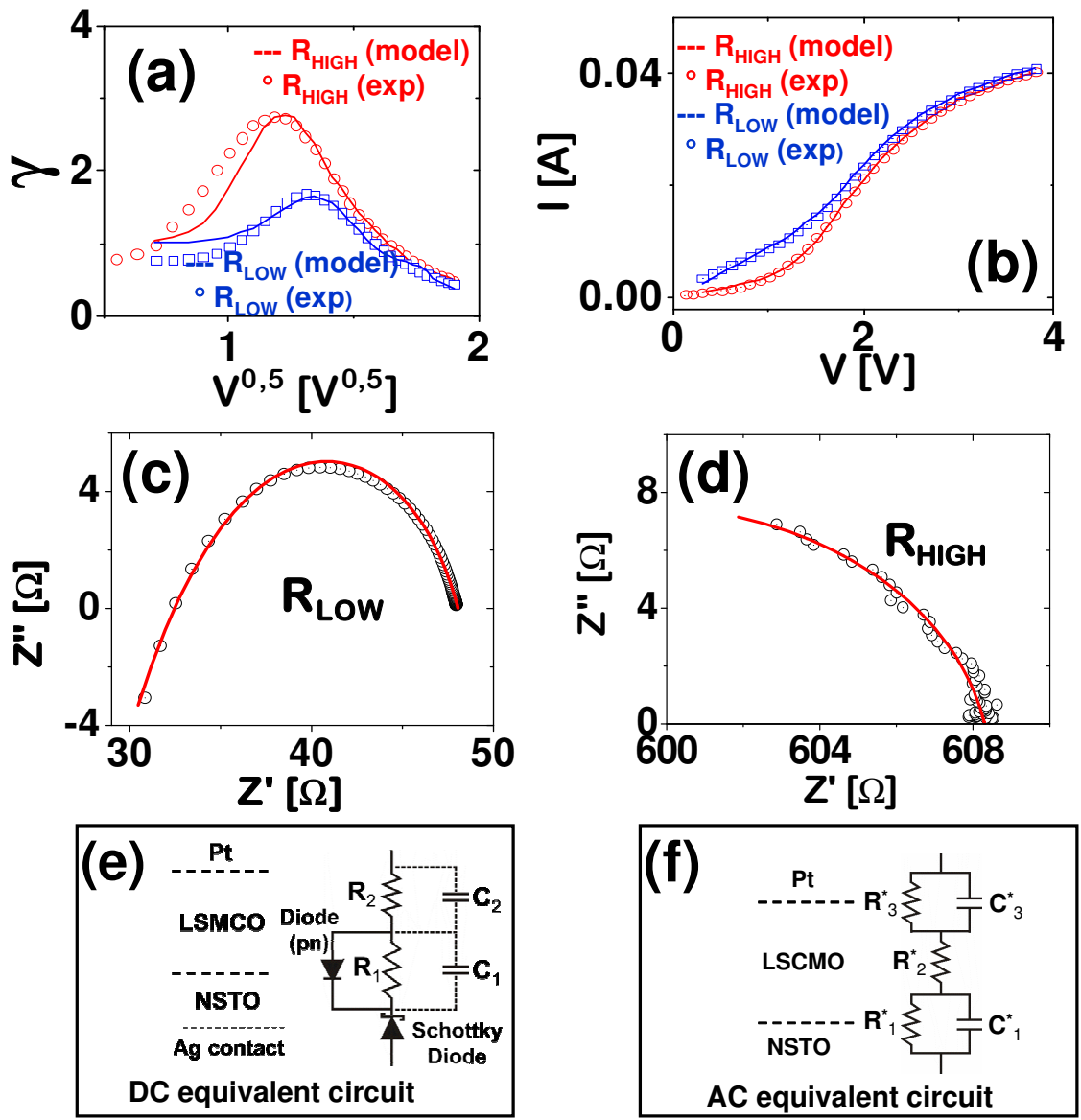

**FIGURE 5**